\documentclass[twocolumn,showpacs,amsmath,amssymb,prd,nofootinbib]{revtex4}
\usepackage{epsfig}
\def\sss{\scriptscriptstyle}
\def\^#1{^{\sss #1}}
\def\_#1{_{\sss #1}}
\def\beq{\begin{equation}}
\def\eeqno#1{\label{#1}\end{equation}}

\def\kms{{\rm km~s^{-1}}}
\def\cmss{{\rm cm~s^{-2}}}
\def\kpc{~{\rm Kpc}}
\def\mpc{~{\rm Mpc}}

\def\msun{M\_{\odot}}
\def\az{a\_{0}}

\def\lsun{L\_{\odot}}
\def\l0{\ell\_{0}}

\def\tp#1{\cdot 10^{#1}}

\def\s{\sigma}
\def\l{\lambda}

\def\a{\alpha}

\def\vv{{\bf v}}

\def\gN{g\_N}

\def\av#1{\langle #1\rangle}
\def\cmss{cm s$^{-2}$}
\def\tot{{2\over 3}}

\def\su{{~\msun/\lsun}}
\def\sigv{\sigma}
\def\sigm{\sigma\_{\small M}}

\begin{document}
\title{MOND in galaxy groups}
\author{Mordehai Milgrom }
\affiliation{Department of Particle Physics and Astrophysics, Weizmann Institute}

\begin{abstract}
Galaxy groups, which have hardly been looked at in MOND, afford probing the acceleration discrepancies in regions of system-parameter space that are not accessible in well-studied galactic systems: galaxies, galaxy clusters, and dwarf spheroidal satellites of galaxies. Groups are typically the size of galaxy-cluster cores, but have masses typically only a few times that of a single galaxy. Accelerations in groups get far below those in galaxies, and far below the MOND acceleration. So much so, that many groups might be affected by the external-field effect, which is unique to MOND, due to background accelerations.
Here, I analyze the MOND dynamics of 53 galaxy groups, recently catalogued in 3 lists. Their Newtonian, K-band, dynamical $M/L$ ratios are a few tens to several hundreds solar units, with $\av{M_d/L\_K}= (56, 25, 30) \su$, respectively for the 3 lists; thus evincing very large acceleration discrepancies. I find here that MOND requires dynamical $M\_M/L\_K$ values of order $1\su$, with  $\av{M\_M/L\_K}=(0.8, 0.56, 1.0) \su$, for the 3 lists, which are in good agreement with population-synthesis stellar values, and with those found in individual galaxies. MOND thus accounts for the observed dynamics in those groups with baryons alone, and no need for dark matter -- an important extension of MOND analysis from galaxies to galactic systems, which, to boot, have characteristic sizes of several hundred \kpc, and accelerations much lower than probed before -- only a few percent of MOND's $\az$. The acceleration discrepancies evinced by these groups thus conform to the deep-MOND prediction: $g\approx (\gN\az)^{1/2}$, down to these very low accelerations ($g$ is the measured, and $\gN$ the baryonic, Newtonian acceleration).
\end{abstract}
\maketitle

\section{\label{introduction} Introduction}
Modified Newtonian dynamics (MOND) \cite{milgrom83,milgrom83a} is an alternative to dark matter striving to account for the acceleration discrepancies in galactic systems, and the Universe at large, by invoking large departures from standard dynamics at low accelerations. MOND --  reviewed in Refs. \cite{fm12,milgrom14c} -- introduces an acceleration constant, $\az$, below which nonrelativistic dynamics is space-time scale invariant.
$\az$ appears in many galactic phenomena whence its value has been determined to be
$\az=1.2\times 10^{-8}$\cmss, with hardy any variation in 30 years, starting with Refs. \cite{milgrom88,bbs} to the recent Ref. \cite{pengfei18}. Intriguingly, this value is near acceleration parameters of cosmological significance, to wit
\beq 2\pi\az \approx ~ a\_{H}(0) ~\equiv ~ c H_0 ~\approx ~ a\_{\Lambda} ~\equiv ~ c^2/\ell\_{\Lambda},
\eeqno{cosmo}
where $H_0$ is the present-day value of the Hubble constant, and $\ell_\Lambda = ( \Lambda/3 )^{-1/2}$ is the de Sitter radius corresponding to the observed value of the cosmological constant $\Lambda$.
\par
MOND has been extensively studied -- arguably quite successfully -- in galaxies of all types: discs, ellipticals, and dwarf spheroidal satellites of more massive galaxies. In galaxy clusters, MOND, while it goes some way in mitigating the acceleration discrepancies, still leaves some discrepancy unexplained.
\par
Galaxy groups -- comprising a few to $\sim$ a hundred galaxies -- lend themselves less easily to reliable dynamical analysis, both in Newtonian dynamics and in MOND. Unlike galaxies and clusters, groups have rather irregular shapes. This makes it more difficult to ascertain that galaxies in their region of the sky are actually group members and gravitationally bound to it. It also make it more difficult to ensure that the group is in virial equilibrium. Thirdly, the irregular shapes do not justify the assumption of system isotropy, an assumption that is made perforce in a dynamical analysis because at present we can only measure the line-of-sight component of member velocities. In addition, for many groups, the velocities of only a small number of members have been measured, greatly exacerbating the above obstacles.
\par
Nonetheless, it is important to study MOND dynamics in groups. First, because they constitute a different type of galactic systems, with possibly different histories.
Secondly, because they probe a region of parameter space hardly accessible in other systems. Studying group dynamics thus enlarges the scope of MOND testing.
\par
For example, the groups studied here have typical baryonic sizes much larger than those of galaxies, but similar to those of cores of galaxy clusters. Their typical baryonic masses are similar to those of (relatively massive) galaxies, and much smaller than those of cluster cores. The characteristic velocities in them, and hence their average gravitational potentials, are also similar to those in galaxies,\footnote{This is a natural consequence of MOND where velocities -- and hence gravitational potentials -- depend only on the mass, but not on size, in the low-acceleration reghime -- a consequence of the scale invariance of the deep-MOND limit.} and much smaller than in cluster cores.\footnote{An observation that is important in the context of attempts to impute the remaining MOND discrepancy in clusters to dependence of $\az$ on the potential \cite{bekenstein11,zf12,milgrom18}.}
\par
Moreover - and as a corollary of the above -- many groups are characterized by very low accelerations, rather lower than what is probed in galaxies using constituent dynamics -- such as rotation curves or velocity dispersions. Only galaxy-galaxy weak lensing has been used to probe MOND to such small accelerations \cite{milgrom13,browuer17}; but this cannot probe individual systems only statistical properties, i.e., very-large-sample averages.
\par
Internal accelerations in groups reach levels of sub-percent to a few percents of $\az$, compared, e.g. with the $\sim 0.1\az$ values reached in galaxies using rotation curves. (And compare, e.g., with Table 1 of Ref. \cite{mm13} for accelerations in dwarf satellites of Andromeda). Groups thus allow us to probe MOND in range of accelerations little explored previously.
\par
To boot, such small accelerations are already at the level expected to be produced in a random position in the Universe, by large scale structure. This may cause observable effect of the internal MOND dynamics through the so-called external-field effect (EFE), which is unique to MOND due to its nonlinear nature \cite{milgrom83} -- a subject of importance in its own right.
\par
Still, probably because of the obstacles mentioned above, the issue has not been visited since 2002, when Ref. \cite{milgrom02a} considered a small sample of 8 relatively near-bye ($D\sim 5 \mpc$) groups from Ref. \cite{tully}. Most of these groups have dynamical times comparable with the Hubble time; so it is questionable whether they are in virial equilibrium, and four of them had no more than 4 members with measured velocities.
Before the advent of Ref. \cite{milgrom02a} there had not been available for study good enough data for individual groups, and analyses such as in Ref. \cite{milgrom98} could analyze only the average properties of a whole sample of not-so-well-defined groups.
\par
Here I take advantage of three rather superior studies of galaxy groups published in recent years, to study MOND dynamics.
They present relevant data for groups in the Hercules-Bootes region \cite{KKK17}, The Leo-Cancer region \cite{KNK15}, and the Bootes-strip region \cite{KKN14}.
They total 53 groups, not all with satisfactory data (for example there are quite a few groups with no more than 4 members with measured velocities, which are not of much use, except perhaps statistically). But there are also many groups that do afford meaningful analysis. These studies also have the advantage that they give galaxy luminosities in the $K$ photometric band, whch is thought to be a better representative of the stellar masses.
\par
In Sec. \ref{method}, I explain the theory underlying the MOND analysis, with the various caveats. Section \ref{analysis} describes the analysis and results. Section \ref{discussion} is a brief discussion of some additional points.
\section{Method and caveats \label{method}}
A general virial theorem for isolated, self-gravitationg systems (ideally of point masses), deep in the MOND regime, was derived in Ref. \cite{milgrom14d} for modified-gravity MOND theories.
It reads
\beq\av{\av{(\vv-\vv_0)^2}}_t=\tot(MG\az)^{1/2}[1-\sum_i
(m_i/M)^{3/2}], \eeqno{nus}
 where $\vv$ is the 3-D velocity,
$\vv_0$ is the center-of-mass velocity,
 $\av{}$  is the mass-weighted average over the constituents, whose
masses are $m_i$, $\av{}_t$ is the long-time average, and $M$ is
the total mass.
In particular, this MOND relation was known to hold in the special cases of the modified-Poisson MOND theory \cite{bm84}, and in Quasilinear MOND \cite{milgrom10a}.
\par
The groups I shall consider here are all very deep in the low-acceleration MOND regime (i.e. with accelerations $\ll\az$). Relation  (\ref{nus}) thus applies in such MOND theories. However, observational limitations force us to use approximations of this relation when applying it to actual systems, as follows: a. {\it Assuming} the system to be quasi stationary, we replace the long-time average with the measured present-day value. b. We cannot measure 3-D velocity dispersions, only line-of sight ones. {\it Assuming} isotrpoy, we replace the former with $\sqrt{3}$ times the latter.
c. Velocity dispersions quoted in the literature are not mass weighted as required in the relation, but based on a limited number of representative members for which line-of-sight velocities are measured and, in the studies we use, are all given equal weight. So we have to make do with these. We thus approximate the left-hand side of relation by $3\sigv^2$, where $\sigv$ is the line-of-sight RMS velosity dispersion.
\par
As to the right-hand side, we are not given the individual masses of all the system members.
We then have to employ one of several approximations. When all the constituent masses are small compared with the total mass, we can neglect the sum in eq. (\ref{nus}), and write
\beq M\approx {81\over 4}\sigv^4(G\az)^{-1}.  \eeqno{nusa}
If the number, $N$, of constituent masses is not small, but they all have comparable masses $m_i\approx M/N$, we can write
\beq M\approx {81\over 4}\sigv^4(G\az)^{-1}(1-N^{-1/2})^{-2}.\eeqno{nusa1}
In the case of one very dominant mass, $M$, with all the rest consisting of `test particles', relation (\ref{nus})
reads $\av{\av{(\vv-\vv_0)^2}}_t=(MG\az)^{1/2}$, and
 gives with our approximations for the left-hand side.
\beq M\approx 9\sigv^4(G\az)^{-1},  \eeqno{nusus}
where $\sigma$ is the mass weighted dispersion of the test particles alone.
\par
Such MOND `virial relations' allow us to determine the total (baryonic) mass in a system from only $\sigv$, unlike the Newtonian relation, which requires also the size of the system. In deep MOND the size disappears because the theory is scale invariant, and under scaling masses and velocities do not change, but sizes do.
\par
More generally, it was shown in Ref. \cite{milgrom14} that a deep-MOND relation of the form $M\approx 9\a\sigv^4(G\az)^{-1}$ -- with $\a$ of order unity, and possibly somewhat system dependent -- follows from only the basic tenets of MOND; so it must be approximately correct in any MOND theory.\footnote{For example, for a central mass $M$ surrounded by an isotropic, deep-MOND, test-particle population, all on circular orbits, any MOND theory predicts a universal rotational speed of $V=(MG\az)^{1/4}$ (the mass-asymptotic-speed relation). For such a system, $\a=1$ for any MOND theory.}
\par
Here I shall use eq.(\ref{nusa}) to estimate the MOND masses of the groups,
because it gives masses that always fall between those of eq. (\ref{nusus}) (larger by a factor 9/4)
and eq. (\ref{nusa1}) (smaller, e.g., by a factor 4 in the case of 4 equal masses).
\par
The number of members with measured velocities may be thought of as one `quality-control' parameter, to which I will add two others.
One is an estimate of the dynamical time, calculated here as the ratio of the harmonic radius, $R_h$ to the one-D velocity dispersion, $\sigv$ -- both given in Refs. \cite{KKK17,KNK15,KKN14}). A necessary (but not sufficient) condition for virial equilibrium is that this quantity is much smaller than the Hubble time. Virialization time may be rather longer than the dynamical time; so requiring only that $\tau_d$ is smaller than the Hubble time is rather mild.
\par
The other parameter to look at is some measure of the internal accelerations in the group $\eta=g/\az\equiv 2\sigv^2/R_h \az$. When this parameters it very low ($\lesssim 0.01$) we have to worry that our assumption of an isolated system is not valid and that the presence of even the background external field due to large-scale structure renders our estimated baryonyonic MOND masses too small (see Sec. \ref{efe}).
\subsection{The external-field effect \label{efe}}
The EFE \cite{milgrom83} is a result of MOND's nonlinearity. In MOND, the internal dynamics of a system (a group in our case) of internal (MOND) acceleration $g$ that is falling in an external gravitational acceleration $g\_{ex}$ are affected by the external field when $g\_{ex}\not\ll g$. There are some anisotropic effects of order unity, but the main effect when $g\_{ex}\gg g$ is to render the internal dynamics Newtonian, but with an effective $G'\approx Gg\_{ex}/g$. The effect was discussed in detail
in \cite{bm84,milgrom10a} for two specific MOND theories. Over the years it has been put to use in different contexts, e.g., recently, in Ref. \cite{mcgaugh16} (a faint dwarf in the field of the Milky Way), Ref. \cite{mm13} (dwarf satellites of Andromeda), Ref. \cite{haghi16} (disc galaxies in the fields of nearby bodies), and Ref. \cite{candlish18} (galxies in clusters).
\par
Because the accelerations in the galaxy groups I study here are only up to a few percent of $\az$ they may allow us to probe the EFE due even to the `random' large-scale-structure acceleration field, whose amplitude is expected to be of the order $10^{-2}\az$. This is, e.g., roughly, the MOND acceleration $35 \mpc$ away from a galaxy cluster
of baryonic mass $10^{14}\msun$. $10^{-2}\az$ is also what is required to accelerate a body to the typical galaxy peculiar velocity of $300\kms$ in the Hubble time.
\par
Reckoning without the EFE when it is present -- i.e., assuming that the system is isolated when in fact, it falls with an external acceleration larger than the internal one -- causes an underestimate of the (baryonic) mass needed to balance inertia.
While it is not possible at present to estimate the field around individual groups, we may see the EFE in action through general trends. For example we may find that groups with particularly small internal accelerations will be found to have unusually small baryonic mass-to-light ratios (compared, e.g., with theoretical values).

\section{Analysis and results \label{analysis}}
I use 3 published samples of groups, totaling 53 objects: 17 groups in the Hercules--Bootes region, cataloged in Ref. \cite{KKK17};
 23 groups in the Leo-Cancer region from
Ref. \cite{KNK15}; and from Ref. \cite{KKN14}, 13 groups in the Bootes-Strip region.
Tables I-III show, respectively, group parameters as deduced in the above references. The distance $D$, the number of members with measured radial velocities, $N\_V$ -- an important quality parameter, the deduced velocity dispersion, $\sigv$, the harmonic radius of the group, $R_h$; and the $K$-band luminosity of the group, $L\_K$. Also, the Newtonian dynamical mass, $M_d$, and the Newtonian dynamical mass-to-light ratio $M_d/L\_K$. No errors are quoted.
\par
For each group I deduce, and give in the tables, the following quantities: an estimate of `the MOND velocity dispersion', $\sigm$, deduced by eq. (\ref{nusa}), assuming the baryonic mass-to-light ratio is $M/L\_K=1$. Next I give the ratio $\sigm/\sigv$. I also give the results of the inverse procedure: I use the published $\sigv$ (uncertain as it must be, especially in cases of a small $N\_V$) to estimate a MOND baryonic mass, $M\_M$, from eq. (\ref{nusa}), and the associated $M\_M/L\_K$.
I also give in the Tables the values of the two other `quality-control' parameters discussed in Sec. \ref{method}.
\par
The Tables speak for themselves, and deserve close scrutiny. Following is a summary of the main relevant points:
\par
1. The Newtonian $M_d/L\_K$ values deduced in Refs. \cite{KKK17,KNK15,KKN14} are typically several tens solar units -- much more than the stellar values expected from stellar-population synthesis models, or from what we observe in individual galaxies, which is $\sim 0.25-1\su$, depending on galaxy type (higher for early types than for late type). This is taken in Newtonian dynamics to bespeak the presence of much dark matter.
\par
2. In contradistinction,
the deduced MOND values are much smaller and fall typically between a fraction of and a few solar units, in agreement with prior knowlede on stellar $M^*/L\_K$ values. The MOND values are baryonic, so would include other baryonic components than stars, such as hot or cold gas.
These values agree well with MOND values deduced for individual disc galaxies from rotation-curve analysis. For example, Ref. \cite{sv98} (Fig. 2) and Ref. \cite{fm12} (Fig. 28) found in this way values of
$M/L\_K\sim 0.25-2\su$, which also agree well with stellar values, and with the trend with galaxy type, obtained in population-synthesis calculations \cite{BdJ01}, as compared e.g., in Ref. \cite{sanders07,fm12}.
So, by and large, MOND accounts for the dynamics within the groups with only baryons as the source of gravity.
\par
3. More quantitatively, for the Hercules--Bootes sample (17 groups), the average $\av{M_d/L\_K}=56 \su$, while with MOND $\av{M\_M/L\_K}=0.79 \su$.
For the Leo-Cancer sample (23 groups): $\av{M_d/L\_K}=30\su$ and $\av{M\_M/L\_K}=0.56\su$.
For the Bootes-Strip sample (13 groups): $\av{M_d/L\_K}=30\su$ and $\av{M\_M/L\_K}=1.0\su$.
\par
4. While the MOND $M/L$ averages are in very good agreement with what has been otherwise known for stellar $M^*/L\_K$ values,
there are quite a few cases with unacceptably small $M/L\_K$ values. However, these occur almost exclusively for groups with `bad' values of the `control parameters', either they have only 3 or 4 measured velocities, or/and they have long dynamical times of order $10^{10}$ years, or/and they have very low, $g<10^{-2}\az$, which makes them highly sucseptable to the EFE.
For example, for the Hercules-Bootes sample, the markedly small $M\_M/L\_K$ values appear only for groups with $N\_V\leq 4$ (and some of these also have small values of $g/\az$, which probably makes them not isolated in the MOND sense). NGC 4866 with a value of $M\_M/L\_K=0.1\su$ has $N\_V=5$, and NGC 4736 ($N\_V=13$) has $M\_M/L\_K=0.18\su$, but $g=4\tp{-3}\az$.
Similarly for the other two samples, with only very few exceptions.
 Since in our analysis, $M\_M/L\_K=1\su$ is equivalent to $\sigm/\sigv=1$, the same trends are seen in the deduced values of $\sigm$.
\par
5.  A closer look at the results raises the possibility that we are seeing signs of the EFE in action, in that for the groups that pass the first two quality control tests, we still encounter low values of $M\_M/L\_K$ correlated with low internal accelerations. For example, in Table I, of the 6 groups with $N\_V>5$, the only one that has a low $M\_M/L\_K=0.18\su$
also has $g=4\tp{-3}\az$. In Table II, 13 groups have $N\_V>5$. Of these, 7 have
$M\_M/L\_K\leq 0.2\su$ (they all have $N\_V\le 10$). All but one have among the smallest $g\le 1.2\tp{-2}\az$.
In table III, 9 groups have $N\_V>5$, 3 of which have unacceptably small $M\_M/L\_K$. They are also the only 3 that have small,
$g\le 7.2\tp{-3}\az$, the other 6 having $g\ge 1.5\tp{-2}\az$.
\par
These indications may be anecdotal at present, but point to a possible, important probe of the background acceleration field and of MOND dynamics.
\par
\begin{figure}[h]
	\centering
\includegraphics[width = 8cm] {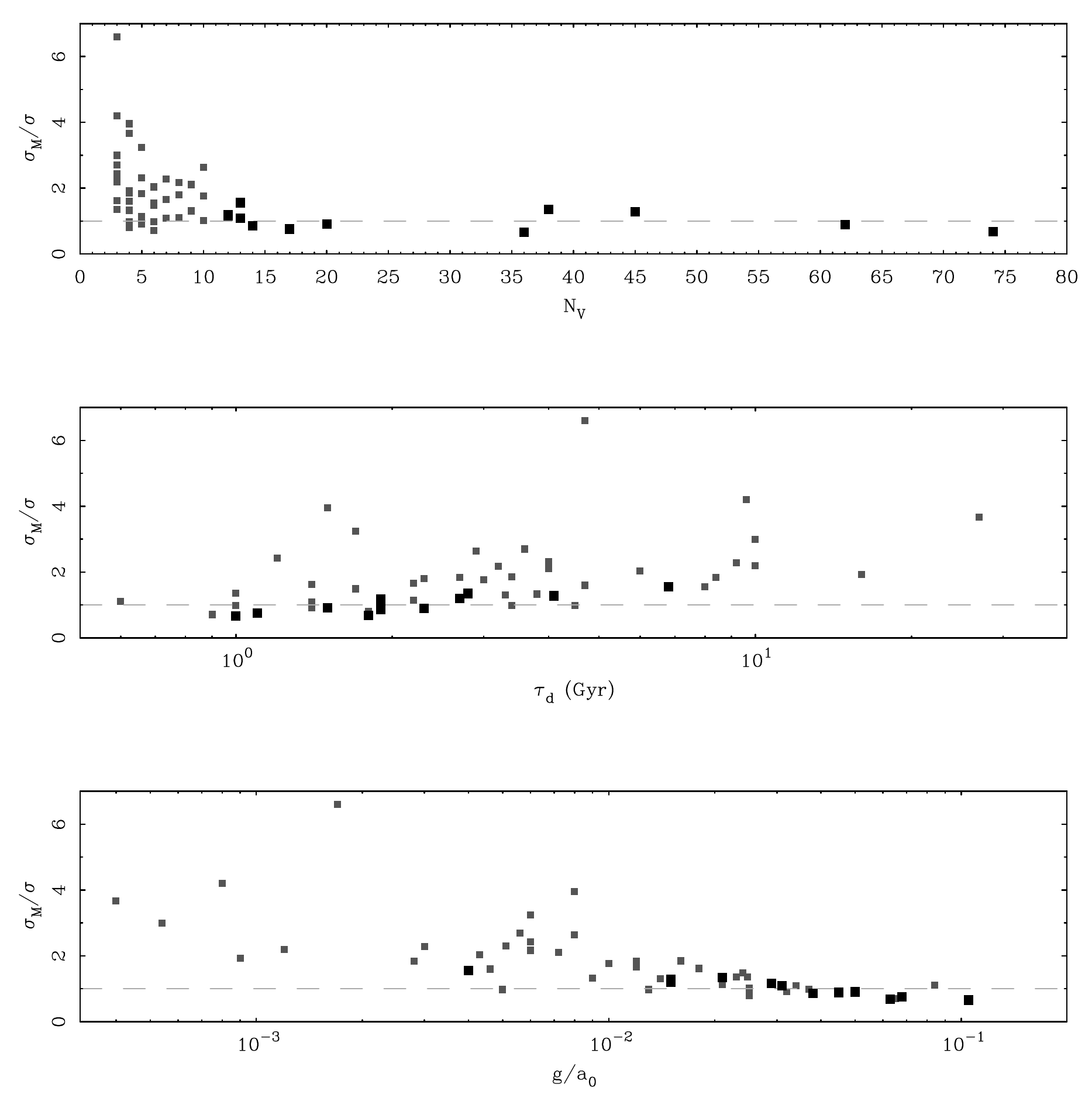}
\caption{The ratios of $\sigm/\sigv$ from Tables I-III,  plotted against the three `quality control' parameters, $N\_V$, the estimated dynamical time, and the estimated internal acceleration (in units of $\az$), in the upper, middle, and lower pannels, respectively. The heavier squares are for groups with more than 10 measured radial velocities. The Figure was prepared by Stacy McGaugh.}		\label{figure1}
\end{figure}

Some of the above points are more clearly seen in Fig. \ref{figure1}, which plots the ratio $\sigm/\sigv$  against $N\_V$, $\tau_d$, and $g$.
\par
For example, we see in the upper panel of Fig. \ref{figure1} that, indeed for groups with larger number of observed velocities the agreement with MOND is very good.
In the lower panel of Fig. \ref{figure1} we see that there is a correlation of $\sigm/\sigv$ with $g$ in the sense that is expected from the presence of an EFE with external accelerations of a few percents of $\az$. However, groups with lower $g$ values also tend to have large dynamical times, which may indicate that they are not virialized yet, leading to smaller-than-virial observed $\sigv$ values. So, this might be underlying the trend of $\sigm/\sigv$ with $g$, not the EFE.

\section{discussion \label{discussion}}
One may wonder why for the groups with a small number of measured velocities we do not also find many with low $\s\_M/\s$ values as well as large ones if the departure from MOND is due to systematics. Note, however, that the complex process of selecting `true' bounded groups in the studies of Refs. \cite{KKK17,KNK15,KKN14} select strongly againts `groups' with high $\s$ values, because of a certain {\it Newtonian} `boundedness' criterion that is applied as a cut. This is particularly true for candidate groups with small number of measured velocities.
\par
Reference \cite{angus08} studied the MOND dynamics of some groups, as part of a more general study of rich groups and clusters, assuming hydrostatic equilibrium of x-ray emitting hot gas in these systems. They do find reduced, albeit remaining acceleration discrepancies in their groups. When comparing to the MOND success in the analysis here, the possibility arises -- discussed in detail in Ref. \cite{milgrom08} -- that the remaining discrepancies in clusters and rich groups has to do with cluster-specific,
yet-undetected baryonic component that is associated with the x-ray-emitting, hot gas, as, by and large, such remaining discrepancies even in MOND do not show up in hot-gas-poor systems.

I thank Stacy McGaugh and Bob Sanders for comments on the manuscript, and Stacy McGaugh for preparing the Figure for me.

\begin{table*}[h]
\caption{The Hercules--Bootes sample (ordered by decreasing number of members with measured velocities). Observed parameters from Ref. \cite{KKK17} with Newtonian parameters as derived there, and MOND-related quantities derived here.
(1) group name; (2) distance in \mpc; (3) number of members with measured line-of-sight velocity; (4) line-of-sight velocity dispersion; (5) harmonic radius in \kpc; (6) $\log\_{10}(L\_K/\lsun)$; (7) dynamical masses $\log\_{10}(M_d/\msun)$; (8) dynamical $M_d/L\_K$ in solar units; (9) MOND velocity diepersions calculated from eq. (\ref{nusa}) assuming baryonic
$M_b/L\_K=1\su$; (10) the ratio $\sigm/\sigv$; (11) MOND masses from $\sigv$ using eq. (\ref{nusa}), in units of $10^{9}\msun$; (12) baryonic, MOND $M\_M/L\_K$ in solar units; (13) a measure of the dynamical time $\tau_d\equiv R_h/\sigv$ in units of $10^{10}$ years;
(14) A measure of the acceleration in the group $g\equiv 2\sigv^2/R_h$ in units of $10^{-2}\az$.}
\medskip
\begin{tabular}{lccccccccccccc}
\hline
\multicolumn{1}{c}{Group}   &  \multicolumn{1}{c}{$D$} & $N\_V$& $\sigv$&   $R_h$&    $\lg L\_K$ &  $\lg M_d$  &  $M_d/L\_K$ & $\sigm$ & $\sigm/\sigv$ & $M\_{M,9}$ &  $M\_M/L\_K$ & $\tau_{d,10}$& \multicolumn{1}{c}{${\small 10^2}g/\az$}\\
\hline \multicolumn{1}{c}{(1)}& \multicolumn{1}{c}{(2)}& (3) &
\multicolumn{1}{c}{(4)} & \multicolumn{1}{c}{(5)}&
\multicolumn{1}{c}{(6)} &\multicolumn{1}{c}{(7)} & (8)
&\multicolumn{1}{c}{(9)} &(10)& (11) & (12) &(13) & (14)
\\ \hline
NGC\,5353 & 35.2 & 62 & 195 & 455 & 12.07 &  13.69 & 42 & 175 & 0.89 & 1829 &  1.56 & 0.23 &  $4.5$\\
NGC\,4736 & 4.5  & 13 &  50 & 338 & 10.64 &  12.33 & 49 & 77 & 1.56  &7.9 &  0.18 & 0.68 &  $0.4$\\
NGC\,5005 & 17.7 & 13 & 114 &  224 & 11.48 &  12.97 & 31& 125 & 1.09 & 211 &  0.70 &  0.19 & $3.1$\\
NGC\,5962 & 33.1 & 8 &  97 &  60  & 11.23 &  13.01 &  60 & 108& 1.11 &115 &  0.65 & 0.06 &  $8.4$\\
NGC\,5582 & 23.1 & 6 & 106 &  93  & 10.60 &  12.44 &  69 & 75 & 0.71 &158 &  3.9 & 0.09 &  $6.5$\\
NGC\,5600 & 25.7 & 6 &  81 &  275 & 10.69 &  12.38 &  49 & 79 & 0.98 &54 &  1.1 & 0.34 &  $1.3$\\
NGC\,4866 & 32.5 & 5 &  58 &  156 & 11.21 &  12.68 &  30 & 107& 1.84 &14 &  0.10 & 0.27 &  $1.2$\\
NGC\,5961 & 31.8 & 5 &  63 &  86  & 10.14 &  12.20 & 115 & 58 & 0.92 &20 &  1.42 & 0.14 &  $2.5$\\
UGC\,10043& 40.4 & 5 &  67 &  65  & 10.37 &  11.88 &  32 & 66 & 0.99 &25 &  1.1  & 0.10 &  $3.7$\\
UGC\,9389 & 32.2 & 4 &  45 &  204 & 9.68  &  12.08 & 251 & 44 & 0.98 &5.1 &  1.1 & 0.45 &  $0.5$ \\
NGC\,5970 & 30.9 & 4 &  92 &  141 & 10.81 &  12.54 &  54 & 85 & 0.92 & 90 &  1.38 &  0.15 & $3.2$\\
NGC\,5117 & 36.3 & 4 &  27 &  424 & 9.97  &  11.95 &  95 & 52 & 1.93 &0.66 &  0.07 &  1.6 & $.09$ \\
NGC\,6181 & 33.9 & 4 &  53 &  196 & 11.06 &  12.14 &  12 & 98 & 1.85 &9.9 &  0.09 & 0.34 &  $1.6$\\
PGC\,55227& 29.4 & 3 &  14 &  17  & 9.21  &  10.05 &   7 & 34 & 2.43 &0.05 &  0.03 & 0.12 &  $0.6$\\
UGC\,10445& 20.6 & 3 &  23 &  230 & 9.92  &  11.60 &  48 & 51 & 2.20 &0.35 &  0.04 & 1.00 &  $0.12$\\
NGC\,5375 & 38.7 & 3 &  47 &  66  & 10.62 &  11.68 &  11 & 76 & 1.62 &6.1 &  0.15 & 0.14 &  $1.8$\\
NGC\,6574 & 29.6 & 3 &  15 &   70 & 11.08 &  10.71 & 0.4 & 99 & 6.6 &0.06 &  $5\times 10^{-4}$ & 0.47 &  $0.17$\\
\hline
\end{tabular}
\end{table*}

\begin{table*}[h]
\caption{The Leo-Cancer sample: The same as Table I with data from Ref. \cite{KNK15}.}
\medskip
\begin{tabular}{lccccccccccccc}\hline
\multicolumn{1}{c}{Group}   &  \multicolumn{1}{c}{$D$} & $N\_V$& $\sigv$&   $R_h$&    $\lg L\_K$ &  $\lg M_d$  &  $M_d/L\_K$ & $\sigm$ & $\sigm/\sigv$ & $M\_{M,9}$ &  $M\_M/L\_K$ & $\tau_{d,10}$& \multicolumn{1}{c}{${\small 10^2}g/\az$}\\
\hline \multicolumn{1}{c}{(1)}& \multicolumn{1}{c}{(2)}& (3) &
\multicolumn{1}{c}{(4)} & \multicolumn{1}{c}{(5)}&
\multicolumn{1}{c}{(6)} &\multicolumn{1}{c}{(7)} & (8)
&\multicolumn{1}{c}{(9)} &(10)& (11) & (12) &(13) & (14)
\\ \hline
NGC\,3607 & 25.0 & 45& 115 & 471 & 11.77 & 13.29 & 33  & 147 & 1.28  & 221  & 0.38 & 0.41 & 1.5 \\
NGC\,3379 & 10.8 & 36& 193& 191 & 11.53 & 13.10 & 37  & 128 & 0.66  & 1755  & 5.2 & 0.10 & 10.5 \\
NGC\,3627 & 10.8 & 20& 136 & 201 & 11.47 &12.96 & 31  & 124 & 0.91  & 433  & 1.5 & 0.15 & 5.0 \\
NGC\,3640 & 27.2 & 14& 134 & 252 & 11.34 & 12.66 & 21  & 115 & 0.86  & 408  & 1.9 & 0.19 & 3.8 \\
NGC\,2962 & 31.6 & 10 & 53 & 161 & 10.99 &11.94& 8.9 & 94 & 1.77 & 10.0  & 0.1 & 0.30 & 1.0 \\
NGC\,3166 & 20.5 & 10& 44 & 126 & 11.36 & 11.97& 4.1 & 116 & 2.64  & 4.7  & 0.02 & 0.29 & 0.8 \\
NGC\,3686 & 21.9 & 10& 91 & 175 & 10.97 & 12.65& 48  & 93 & 1.02  &  87 & 0.93 & 0.19 & 2.5 \\
NGC\,2775 &26.9  & 9&  89 & 296 & 11.37 & 12.99 & 42  & 117 & 1.31 & 79  & 0.34 & 0.33 & 1.4 \\
NGC\,2648 & 36.0 & 8& 55 & 128 & 11.09 & 11.98& 7.8 & 99 & 1.80 & 11.6  & 0.09 & 0.23 & 1.2 \\
NGC\,3338 & 20.1 & 7& 50 & 112 & 10.77 & 11.05 & 1.9 & 83 & 1.66  &  7.9 & 0.13 & 0.22& 1.2 \\
NGC\,2894 & 39.6 & 7& 50 & 458 & 11.32 & 12.23 & 8.1 & 114 & 2.28 & 7.9  & 0.04 & 0.92 & 0.3 \\
NGC\,2967 & 35.8 & 6& 62 & 507 & 11.03 & 12.75 & 52  & 96 & 1.55 & 18.7  & 0.17 & 0.80 & 0.4 \\
NGC\,3227 & 25.7 & 6& 74 & 128 & 11.27 & 12.50 & 17  & 110 & 1.49  & 38  & 0.2 & 0.17 & 2.4 \\
NGC\,3023 & 28.8 & 5& 21 & 35 & 10.44 & 11.40  & 9.1 & 68 & 3.24  & 0.24  & 0.01 & 0.17 & 0.60 \\
NGC\,3626 & 25.6 & 5& 86 & 187 & 11.06 & 12.75   & 49  & 98 & 1.14  & 69  & 0.6 & 0.22 & 2.1 \\
NGC\,3810 & 17.7 & 5& 43 & 360 & 10.67 & 12.12 & 28  & 78 & 1.84  & 4.3  & 0.09 & 0.84 & 0.28 \\
NGC\,3423 & 23.1 & 4& 21 & 570 & 10.64 & 12.14 & 32  & 77 & 3.67  & 0.25  & 0.005 & 2.7 & 0.04 \\
UGC\,5228 & 32.7 & 4& 40 & 188 & 10.31 & 11.90 & 39  & 64 & 1.60  & 3.2  & 0.16 & 0.47 & 0.46 \\
UGC\,5376 & 45.3 & 4& 66 & 253 & 10.87 & 12.23 & 23  & 88 & 1.33  & 24  & 0.32 & 0.38 & 0.9 \\
NGC\,3521 & 10.7 & 3& 37 & 132 & 11.10 & 12.52 & 26  & 100 & 2.70  & 2.4  & 0.02 & 0.36 & 0.56 \\
NGC\,3020 & 30.2 & 3& 45 & 44 & 10.24 & 11.53  & 19  & 61 & 1.36  & 5.2  & 0.30 & 0.10 & 2.48 \\
NGC\,3049 & 30.2 & 3& 15 & 144 & 10.29 & 11.31 & 10  & 63 & 4.20  & 0.064  & 0.003 & 0.96 & 0.08 \\
NGC\,3596 & 14.0 & 3& 42 & 41 & 10.13 & 11.43  & 20  & 57 & 1.36  & 3.9  & 0.37 & 0.1 & 2.3 \\
\hline
\end{tabular}
\end{table*}

\begin{table*}[h]
\caption{The Bootes-Strip sample: The same as Table I with data from Ref. \cite{KKN14}.}
\medskip
\begin{tabular}{lccccccccccccc}\hline
\multicolumn{1}{c}{Group}   &  \multicolumn{1}{c}{$D$} & $N\_V$& $\sigv$&   $R_h$&    $\lg L\_K$ &  $\lg M_d$  &  $M_d/L\_K$ & $\sigm$ & $\sigm/\sigv$ & $M\_{M,9}$ &  $M\_M/L\_K$ & $\tau_{d,10}$& \multicolumn{1}{c}{${\small 10^2}g/\az$}\\
\hline \multicolumn{1}{c}{(1)}& \multicolumn{1}{c}{(2)}& (3) &
\multicolumn{1}{c}{(4)} & \multicolumn{1}{c}{(5)}&
\multicolumn{1}{c}{(6)} &\multicolumn{1}{c}{(7)} & (8)
&\multicolumn{1}{c}{(9)} &(10)& (11) & (12) &(13) & (14)
\\ \hline
N5846  & 26.4& 74 &  228  & 415 & 11.84 & 13.66&  66  & 155 & 0.68 & 3418  & 4.9  & 0.18  & 6.3 \\
N5746  & 25.6& 38 &  107  & 296 & 11.74 & 13.24&  32  & 145 & 1.35 & 165  & 0.30  & 0.28  & 2.1 \\
N5363  & 17.5& 17 &  144  & 165 & 11.25 & 12.80&  35  & 109 & 0.76 & 544  & 3.1  & 0.11  &  6.8 \\
N5566  & 22.3& 12 &  103  & 196 & 11.35 & 12.81&  29  & 115 & 1.17 & 142  & 0.63  & 0.19  & 2.9 \\
N5638  & 25.8& 12 &   74  & 203 & 10.90 & 12.21&  20  & 89  & 1.20 & 38.0  & 0.48  & 0.27  & 1.5 \\
N5838  & 25.0&  9 &   53  & 210 & 11.31 & 12.29&  9.5 & 113 & 2.11 & 10.0  & 0.05  & 0.40  & 0.72  \\
N4900  & 23.6&  8 &   36  & 116 & 10.67 & 11.95&  19  &  78 & 2.17 & 2.1  & .05  & 0.32  & 0.60 \\
N5775  & 20.1&  7 &   87  & 120 & 11.01 & 12.66&  45  & 95  & 1.09 & 72  & 0.71  & 0.14  & 3.4 \\
N5792  & 25.4& 6  &   48  & 290 & 11.06 & 12.00&  8.7  & 98 & 2.04 & 6.7  & 0.06  & 0.60  & 0.43 \\
N5248  & 17.1&  5 &   38  & 151 & 10.88 & 12.06&  15  & 88  & 2.31 & 2.6  & .03  & 0.40  & 0.51 \\
IC1048 & 26.4&  4 &   83  & 150 & 10.38 & 12.32&  87  & 66  & 0.80 & 60.0  & 2.5  & 0.18  & 2.5 \\
N5506  & 23.8&  4 &   23  &  35 & 10.94 & 11.07&  1.4  & 91 & 3.96 & 0.35  & $4\tp{-3}$  & 0.15  & 0.80 \\
P51971 & 20.6& 3  &   10  & 100 &  9.04 & 10.35&  20  & 30  & 3.00 & 0.013  & 0.012  & 1.0  & .054 \\

\hline
\end{tabular}
\end{table*}

\end{document}